\documentstyle[multicol,prl,psfig,aps]{revtex}
\begin{document}

\title{
Electronic properties of the 1D
Frenkel-Kontorova model}

\author{Peiqing Tong$^{1,3}$, Baowen Li$^2$, and Bambi Hu$^{3,4}$
}
\address{
$^1$  Department of Physics, Nanjing Normal University, Nanjing, Jiangsu,
210097, China\\
$^2$ Department of Physics, National University of Singapore, 117542 Singapore\\
$^3$Department of Physics and Centre of Nonlinear Studies, Hong Kong
Baptist University, Hong Kong, China,\\
$^4$Department of Physics, University of Houston, Houston, TX
77204-506}
\date{Phys. Rev. Lett. {\bf 88}, 046804 (2002)}
\maketitle

\begin{abstract}
The energy spectra and quantum diffusion of an electron in
a 1D incommensurate Frenkel-Kontorova (FK) model
are studied numerically. We found that the spectral and dynamical
properties of electron display quite different behaviors in invariance
circle regime and in Cantorus regime. In the former case, it is similar
to that of the Harper model, whereas in the latter case, it is similar to that
of the Fibonacci model. The relationship between spectral and
transport properties is discussed.
\end{abstract}

\pacs{61.44.Fw, 05.45+Mt, 73.20.Dx, 05.60.Gg}

\begin{multicols}{2}

The Frenkel-Kontorova model describes a one dimensional chain of
atoms/particles with harmonic nearest neighbor interaction placed
in a periodic potential. It is a widely used model in condensed
matter physics and nonlinear dynamics\cite{FK}. For instance, it
has been used to model crystal dislocations\cite{Naba67},
epitaxial monolayers on the crystal surface\cite{Ying71}, ionic
conductors and glassy materials\cite{Glassy}, electron in a quasi
1D metal below the Peierls transition\cite{sc85}, charge density
waves\cite{CDW}, Josephson junctions chains\cite{JJ}, and dry
friction\cite{Friction}. More recently, this model has been
employed to study transport properties of vortices in easy flow
channels\cite{Besseling} and strain-mediated interaction of
vacancy lines in a pseudomorphic adsorbate system\cite{Erwin}. Due
to the competition between the two length scales, the spring
length and the period of the on-site potential, the FK model
exhibits a wealth of interesting and complex phenomena (see
\cite{Braun} and the reference therein).

One of the most striking features of the FK model is the so-called
transition by breaking of analyticity. It is shown by
Aubry\cite{au83} that there exist two different ground state
configurations for an incommensurate chain. The transition from
one configuration to another occurs when one changes the coupling
constant $K$ (see Eq.(\ref{fkp})). These two incommensurate
configurations correspond to invariance circle and Cantorus of the
standard map\cite{ch79}, respectively.  This transition is still
discernible in a quantum FK model\cite{hl98}.

Although extensive studies have been done since its introduction,
the FK model continues to attract active interests from different
fields. Recent studies are concentrated on phonon modes, because
they are responsible for the heat conduction along the
chain\cite{Phonon,HLZ}. It is found that the on-site potential
breaks the conservation of momentum, and makes the heat conduction
in 1D FK model obey the Fourier law\cite{HLZ}.

However, the electronic property of the 1D FK model is still unknown up to
now. This topic is  important from both fundamental and application points of view, because
incommensurate and quasiperiodic structures appear in many physical
systems such as quasicrystals,
two-dimensional electron systems, magnetic superlattices, charge-density
waves, organic conductors and various atomic monolayers absorbed on
crystalline substrates. As mentioned before, the FK model has been very successful applied in these systems. It is the purpose of this Letter to study this topic.

The electron in a 1D FK chain obeys the equation
\begin{equation}
t_0(\psi_{n+1}+\psi_{n-1})+V_n\psi_n=E\psi_n,
\label{ham}
\end{equation}
where $t_0$ is a nearest-neighbor hopping integral which is set to 1
in this Letter, $\psi_n$ is the amplitude of wave function at the
$n$th site and $E$ the eigenenergy of electron.
The on-site potential, $V_n=\lambda\cos(x_n^0)$, is controlled by
parameters $\lambda$ and $x_n^0$.  $\{ x_n^0\}$ is the configuration of an
incommensurate ground state of the FK model, namely,
$\{x^0_n\}$ minimizes the functional
\begin{equation}
U=\sum_n
\frac{1}{2}(x_{n+1}-x_n-a)^2+K(1-\cos(x_n)),
\label{fkp}
\end{equation}
where $K$ is a coupling constant, and
$a$ the equilibrium
distance between consecutive atoms.  $\{x_n^0\}$ is determined by an
adjustable coupling constant $K$.

In contrast to the Harper model and the Fibonacci model that have
been often used to study electron in incommensurate systems (see
review articles\cite{Sokoloff,Harper2}, Ref. \cite{Fibonacci}, and
the references therein), the FK model has two control parameters,
$\lambda$ and $K$, and is thus more general. As we shall see below
that the two control parameters not only make the problem more
complicated but also enrich the physics of this model. Most
electron properties shown in the Harper model and the Fibonacci
model can be recovered in the FK model.

Any change in $\lambda$ and $K$ will alter on-site potential
$V_n$, thus change the electron properties of the system. As it is
well known that for each irrational number $a/2\pi$ there exists a
critical value $K_c$ separating the two configurations of ground
state. $K_c=0.9716354\cdots$ corresponds to the most irrational
number, golden mean value $a/2\pi=(\sqrt{5}-1)/2$. In this Letter,
we restrict ourselves to this particular value in numerical
calculations as it is the mostly used one in the community.

In order to study electron energy spectra, we first
obtain ground state configuration for $N$ atoms
by gradient method \cite{au83} for fixed boundaries, i.e.,
$x_0\equiv0$ and $x_N=2\pi N$.
$a/2\pi=(\sqrt{5}-1)/2$ is approximated by a
convergent series of truncated fraction: $F_n/F_{n+1}$
($n=1,2\cdots\cdots$), where $\{ F_n\}$ is a Fibonacci sequence. The
number of atoms is chosen as $N=F_n$. The electron eigenenergies are
obtained numerically by the transfer matrix method.
The transfer matrix ${\bf T}_N(E)$ is
\begin{equation}
{\bf T}_N(E)=\prod\limits_{n=0}^{N}{\bf T}(n,E),
\end{equation}
where
\begin{equation}
 {\bf T}(n,E)=\left(
{E-\lambda\cos(x_n^0) \qquad -1}\atop {1 \hspace{2.25cm} 0}
\right).
\end{equation}

\begin{figure}
\centerline{\psfig{figure=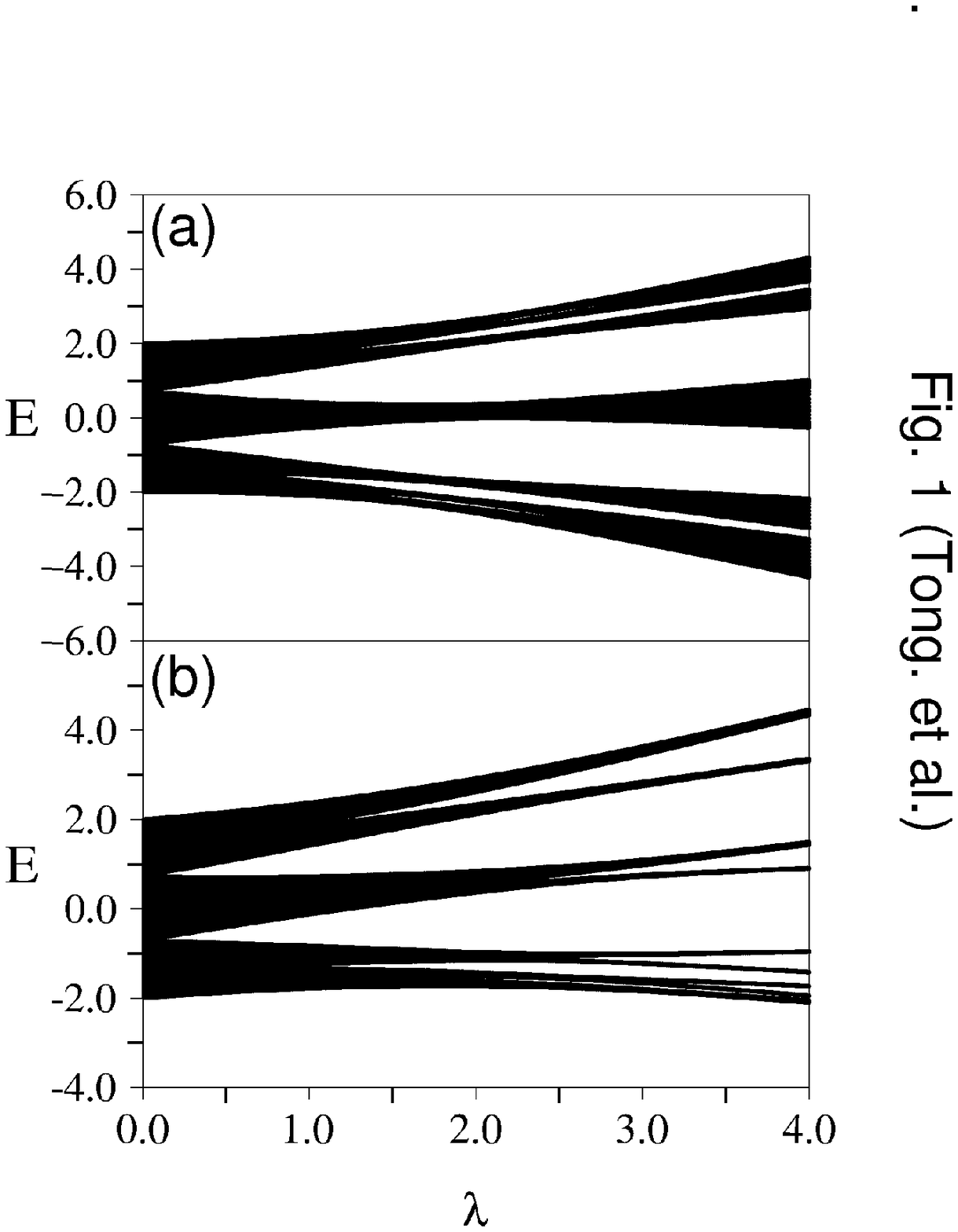,height=9cm,angle=180}}
\vspace{-1.cm} \narrowtext \caption{Energy spectra versus
$\lambda$ for electron in the FK chains with different values of
$K$. (a), $K=0.4<K_c$; (b), $K=1.6>K_c$. The chain length is
$N=377$.} \label{fig1}
\end{figure}
The allowed energies of electron satisfy condition $|Tr{\bf
T}_N(E)|\leq 2$. Fig. \ref{fig1} illustrates the spectra as functions of
$\lambda$ for different values of $K$. From Fig.
\ref{fig1}, we can see that for $K\leq K_c$, the energy spectra (see Fig
\ref{fig1}a) is similar to that of the Harper model, namely, the band splits
into subbands as $\lambda$ is increased from $0$. As $\lambda$ becomes larger
than a critical value, energy levels tend to repel each other.
There are however differences from the Harper
model. For instance, the Harper model has a good
symmetry, such as self-duality, and the spectrum is symmetric about
$E=0$. All eigenstates are extended when $\lambda <\lambda_c(=2)$, and localized when $\lambda >\lambda_c$. Our model does not have self-duality and the
spectrum is asymmetric. It is
known \cite{Sokoloff} that for a non-self-duality system, there exits a critical
parameter above that all eigenstates are localized. Below this
critical value, extended, critical, localized
states, and mobility edges coexist.

To study the eigenstates quantitatively, the Thouless
exponent ($\gamma(E_i)$) and participation ratios (PR) are calculated for every eigenstate. The Thouless exponent is given by
\begin{equation}
\gamma(E_i)=\int{\ln|E_i-E_j|\rho(E_j)dE_j}=\frac{1}{N}\sum_{j\neq i}\ln
|E_i-E_j| ,
\end{equation}
where $\rho(E)$ is the density of states,
and the participation ratio is,
\begin{equation}
PR=\frac{(\sum_n \psi_n^2)^2}{N\sum_n \psi_n^4}.
\end{equation}
\begin{figure}
\centerline{\psfig{figure=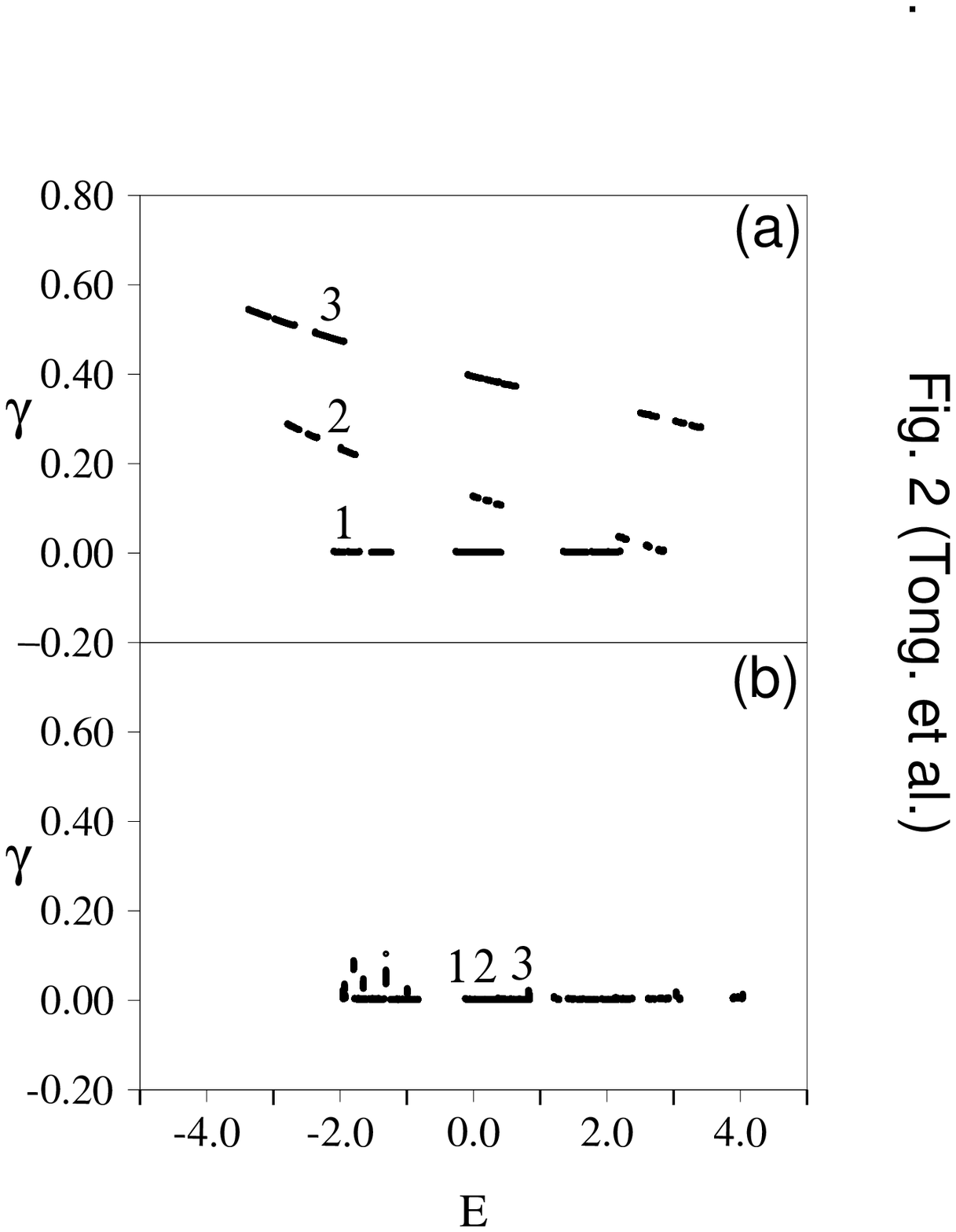,height=9cm,angle=180}}
\vspace{0cm} \narrowtext \caption{The Thouless exponent
$\gamma(E)$ for different values of $K$. (a), $K=0.4$, curves 1, 2
and 3 correspond to $\lambda=1.0$, $2.3$ and $3.0$, respectively;
(b), $K=1.6$, curves 1, 2 and 3 correspond to $\lambda=1.0$, $2.0$
and $3.5$, respectively. The chain length is $N=4181$.}
\label{fig2}
\end{figure}
The Thouless exponent is proportional to the inverse of
localization length i.e. $\gamma \sim 1/\xi$. If $\gamma$ is about
order of $1/N$ for finite chain of length $N$, then the
eigenstates are extended or critical. Otherwise the eigenstates
are localized. In Fig. \ref{fig2}, we plot $\gamma$ as a function
of eigenstate for different $\lambda$. Fig. \ref{fig2}a is for
$K=0.4$. It tells us that for small $\lambda$, all states are
extended and critical (correspond to small $\gamma$ in Fig.
\ref{fig2}a). As $\lambda$ is increased, some eigenstates become
localized. For $\lambda>\lambda_c$ all eigenstates are localized.
Of course, $\lambda_c$ depends on $K$\cite{sc85}. The minimum
Thouless exponent $\gamma_{min}$ and maximum PR are calculated as
a function of $\lambda$ so as to find the critical value of
$\lambda_c$ for a fixed $K$. In Fig. \ref{fig3}, we plot
$\gamma_{min}$ and $PR_{max}$ versus $\lambda$ for $K=0.4$, $0.6$,
$K_c$, and $1.6$. From these curves, we can easily obtain critical
values $\lambda_c\approx 2.3$, $2.46$ and $2.96$ for $K=0.4$,
$0.6$ and $K_c$, respectively.

In the case of $K>K_c$, i.e., the ground state configuration of atoms
corresponds to Cantorus, both spectra and eigenstates
are quite different from that case of $K\leq K_c$ (see Fig. \ref{fig2}b).  In
this case, no critical value has been found  (see curve 4 and curve 2 in Fig. (3a) and (3b), respectively.).
All eigenstates are
critical. This is similar to that case of
quasiperiodic systems such as the Fibonacci chain\cite{Fibonacci}.

\begin{figure}
\centerline{\psfig{figure=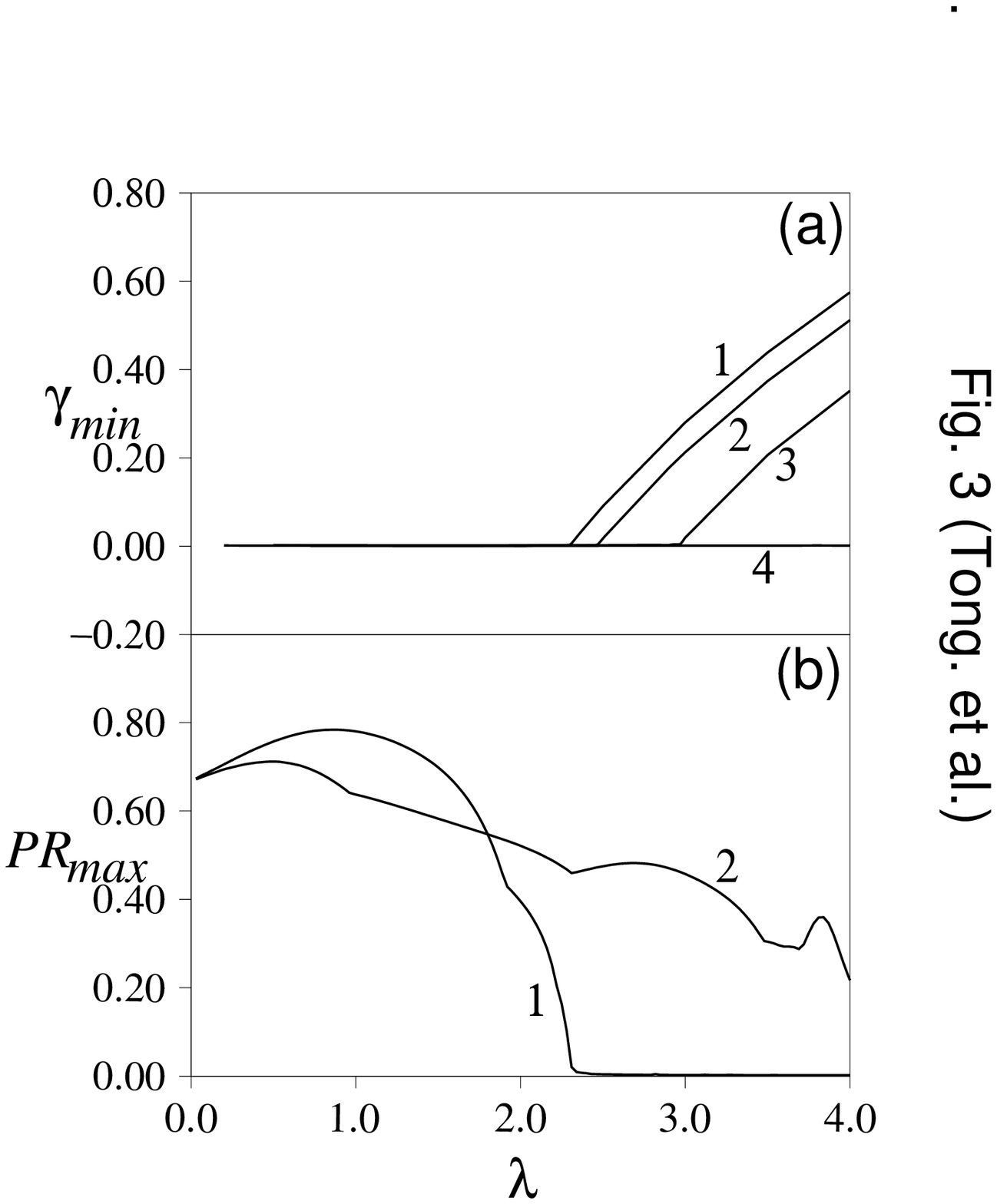,height=9cm,angle=180}}
\vspace{0.5cm} \caption{The minimum Thouless exponent
$\gamma_{min}$ (a)  and the maximum participation ratio $PR_{max}$
(b) as functions of $\lambda$ for electron in the FK chains with
different parameter $K$. Curves 1, 2, 3 and 4 in (a) correspond to
$K=0.4$, $0.6$, $K_c$ and $1.6$. Curves 1 and 2 in (b) correspond
to $K=0.4$ and $1.6$. The results of $\gamma_{min}$ and $PR_{max}$
are obtained for finite FK chains of length $N=4181$ and $N=1597$,
respectively.} \label{fig3}
\end{figure}
To investigate quantum dynamical behaviors,
the time evolution of a wavepacket in the system described by
Eq. (\ref{ham}) is
calculated numerically. The wavepacket is localized initially at the
centre of the chain.
The time evolution is described by a time-dependent Schr\"{o}dinger
equation

\begin{equation}
i\frac{d\psi_n}{dt}=\psi_{n+1}+\psi_{n-1}+\lambda\cos(x_n^0)\psi_n.
\label{dyn}
\end{equation}
The variance of the wavepacket is
\begin{equation}
\sigma^2(t)=\sum_{n=1}^{N} (n-\bar{n})^2 | \psi_n(t)|^2.
\end{equation}
It can be calculated numerically by integrating the
Schr\"{o}dinger Eq. (\ref{dyn}) for a chain of length $N$ with
fixed boundaries $\psi_0=\psi_{N+1}=0$. In our calculations, the
fourth-order Runge-Kutta method with time step $\delta t=0.01$ was
used. The equilibrium ground state positions of $N$ atoms in the
FK chain are obtained by the gradient method with the same
boundary conditions. The dynamical exponent is defined by,
$\sigma^2\sim t^{2\Delta}.$
\begin{figure}
\centerline{\psfig{figure=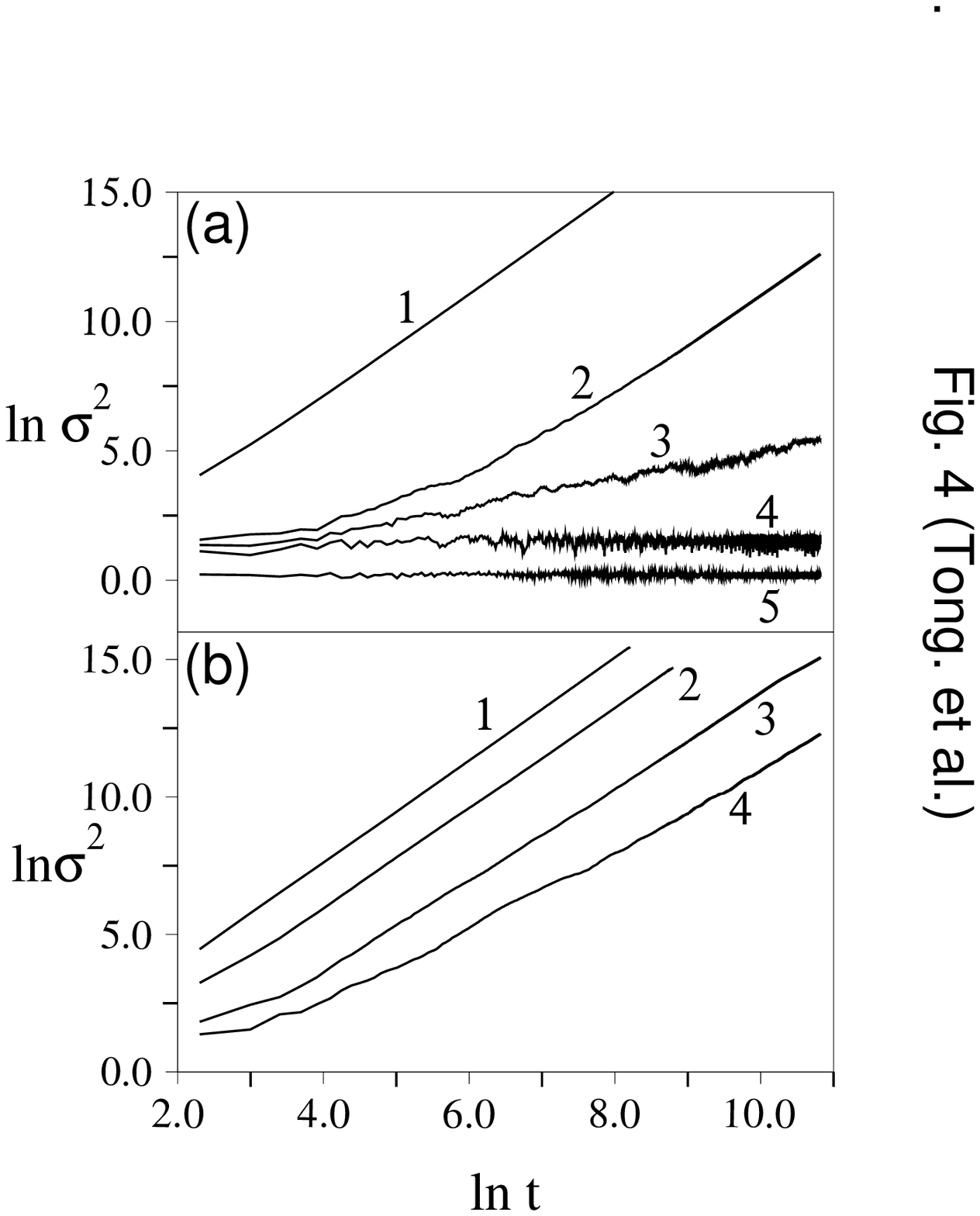,height=9cm,angle=180}}
\vspace{0.5cm} \narrowtext \caption{The variance $\sigma^2$ of a
wavepacket for electrons in the FK chains with different values of
$K$. (a), $K=0.4$, curves 1, 2, 3, 4 and 5 correspond to
$\lambda=1.0$, $2.2$, $2.3$, $2.4$, and $3.0$, respectively, and
$\lambda_c\approx 2.3$; (b), $K=1.6$, curves 1, 2, 3 and 4
correspond to $\lambda=1.0$, $2.0$, $3.0$ and $3.5$, respectively.
The chain length is $N=10946$.} \label{fig4}
\end{figure}
In Fig. \ref{fig4}, we plot
$\sigma^2(t)$ for several values of $\lambda$. Figs.  \ref{fig4}a and
\ref{fig4}b
correspond to $K=0.4$ and $1.6$,
respectively. For $K<K_c$,
the time evolution of electron is also similar to
that of the Harper model,
that is $\sigma^2\sim t^2$ and $t^0$ for $\lambda<\lambda_c$
($\lambda=1.0, 2.2$) and
$\lambda>\lambda_c$ ($\lambda=2.4$), respectively.
The unbounded diffusion in the regime of $\lambda <\lambda_c$ is caused by the existence of extended and critical states as discussed above.
However, at  $\lambda=\lambda_c$, the system displays anomalous diffusion
behaviors,
and the dynamical exponent depends on $K$, and
$\Delta \approx 1/4$ for $K=0.4$ (see Fig. \ref{fig4}a). It
will be of
great
interest
to connect the exponent of anomalous diffusion with the multifractal
dimension of critical eigenstates. For
$K>K_c$, the time behaviors of wavepacket are
similar to that of quasiperiodic systems, i.e. the dynamical exponent
$\Delta$ depends on $\lambda$.

\begin{figure}
\centerline{\psfig{figure=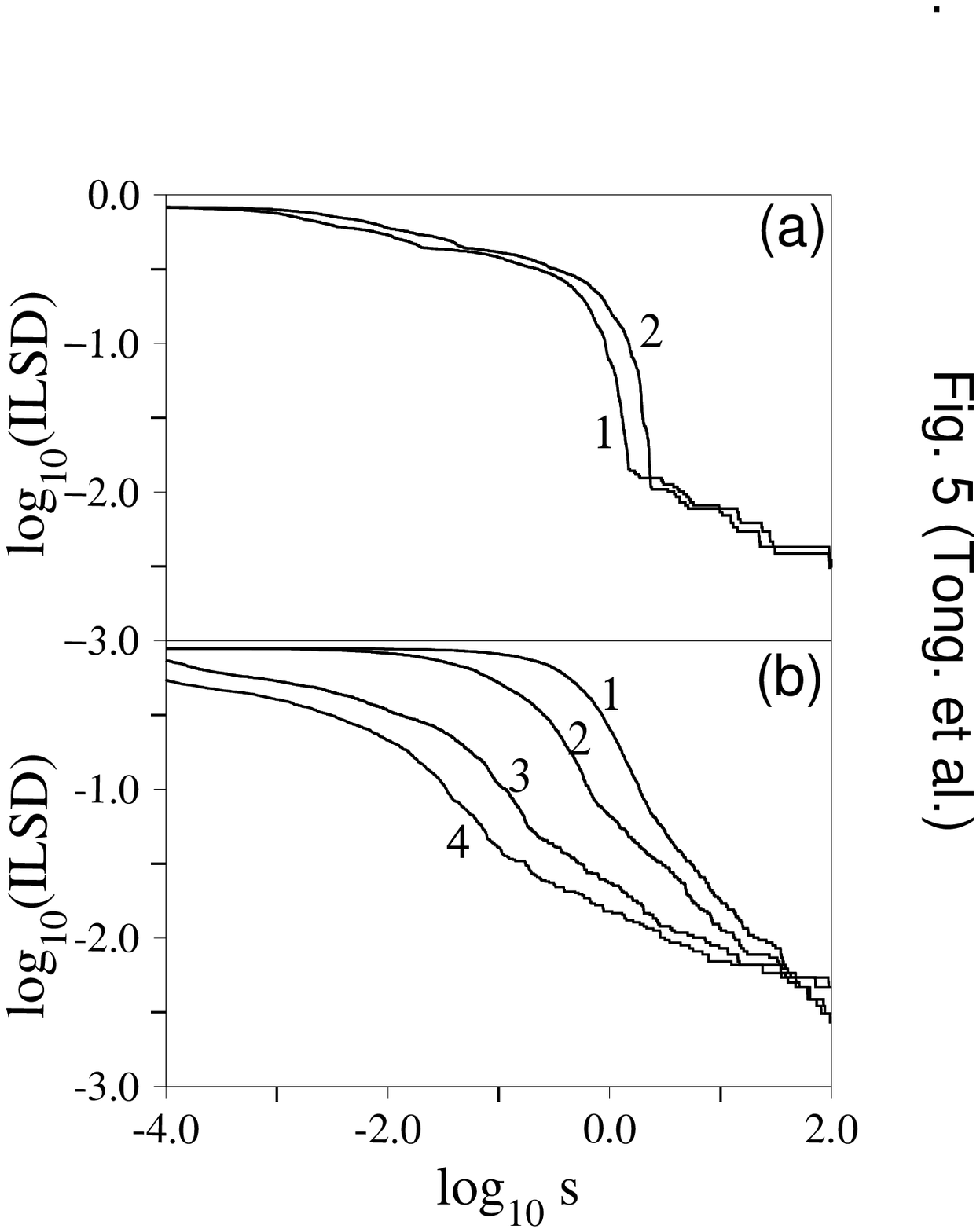,height=9cm,angle=180}}
\vspace{0.5cm} \narrowtext \caption{Integrated level-spacing
distribution $p_{int}(s)$ for different values of $K$. (a), Curve
1 corresponds to $K=0.4$ and $\lambda_c=2.3$. Curve 2 corresponds
to $K=0.6$ and $\lambda_c=2.46$; (b), $K=1.6$, curves 1, 2, 3 and
4 correspond to $\lambda=1.0$, $2.0$, $3.0$ and $3.5$,
respectively. The chain length is $N=2584$.} \label{fig5}
\end{figure}

Now we turn to level statistics of the system and its
relationship with the
dynamical exponent. Geisel {\sl et. al.} \cite{ge95} observed
that for bounded uncountable set of levels, it is possible to count the
number of energy gaps larger than $s$ and to calculate
the integrated
level-spacing distribution (ILSD) defined by
$p_{int}(s)\equiv \int_{s}^{\infty}p(s\prime) ds\prime.$
The derivative of ILSD $p(s)=-dp_{int}/ds$ gives probability
distribution of level spacings. In Fig. \ref{fig5}a we show ILSD at
$\lambda=\lambda_c$ for $K=0.4$ and $0.6$.
It can be seen that each distribution
consists of two parts: exponentially decay part for small $s$ and
power-law decay part for large $s$. It is well known that the localization of eigenstates results in the Poisson distribution in energy level spacing statistics\cite{ma86}. It is thus reasonable to
attribute the exponentially decay to the
localized states, and the power-law like decay to
the critical states, as is the case in the Harper model and
the quasiperiodic model.  Fig. \ref{fig5}b shows ILSD for several
values of $\lambda$ with $K=1.6$. Each
distribution is very
similar to that of the Harper model at $\lambda_c$ and that of
quasiperiodic models. Unfortunately, the power-law like behavior of ILDS
is not very significant for large $\lambda$ due to limited bin sizes.

The exponent $\beta$ for level-spacing distribution defined by
$p(s)\sim s^{-\beta}$ can be obtained by best fit power-law part of
ILSD curves. We find that the relation
\begin{equation}
\Delta=1-\beta,
\label{exp}
\end{equation}
for the Harper model at $\lambda_c$ is also true for our model in
the regime of $K<K_c$, but not for the regime of $K>K_c$.

In summary, we have studied spectral and dynamical properties
of an electron in incommensurate FK chains. The system shows rich
phenomenon. For $K<K_c$, i.e., the ground  state configuration
corresponds to invariance circle, there exists
a critical value $\lambda_c$ above which all eigenstates are localized.
Below $\lambda_c$, extended,
critical and localized states coexist.  The critical value $\lambda_c$ and
dynamical exponent $\Delta$ at $\lambda_c$
depend on $K$. The relation (\ref{exp}) holds for power-law
part of ILSD.  On the other hand, for the case of $K>K_c$, i.e., the
ground state configuration
corresponds to Cantorus, all electron
eigenstates are critical and $\Delta$ depends on $\lambda$.

P.T. was supported in part by the National Nature Science
Foundation of China under Grant No. 10175035. B.L. was supported
in part by Academic Research Fund of National University of
Singapore.

\end{multicols}
\end{document}